\documentclass[runningheads]{llncs}
\usepackage{graphicx}

\usepackage[usenames, dvipsnames]{color}

\newcommand{\tabincell}[2]{\begin{tabular}{@{}#1@{}}#2\end{tabular}}

\begin{document}
\title{
3DFPN-HS$^2$: 3D Feature Pyramid Network Based High Sensitivity and Specificity Pulmonary Nodule Detection
}

\titlerunning{3DFPN-HS$^2$ based Pulmonary Nodule Detection}

\author{Jingya Liu\inst{1} \and
Liangliang Cao\inst{2,3} \and
Oguz Akin \inst{4}\and
Yingli Tian\inst{1,\thanks{Corresponding author. Email: ytian@ccny.cuny.edu}}}

\institute{The City College of New York, New York, NY 10031\and
UMass CICS, Amherst, MA 01002\and 
Google AI, New York, NY 10011\and
Memorial Sloan-Kettering Cancer Center, New York, NY, 10065} 


%
\authorrunning{J. Liu et al.}
%

%
\maketitle              

\begin{abstract} 

Accurate detection of pulmonary nodules with high sensitivity and specificity is essential for automatic lung cancer diagnosis from CT scans. Although many deep learning-based algorithms make great progress for improving the accuracy of nodule detection, the high false positive rate is still a challenging problem which limited the automatic diagnosis in routine clinical practice. In this paper, we propose a novel pulmonary nodule detection framework based on a 3D Feature Pyramid Network (3DFPN) to improve the sensitivity of nodule detection by employing multi-scale features to increase the resolution of nodules, as well as a parallel top-down path to transit the high-level semantic features to complement low-level general features. Furthermore, a High Sensitivity and Specificity (HS$^2$) network is introduced to eliminate the falsely detected nodule candidates by tracking the appearance changes in continuous CT slices of each nodule candidate. The proposed framework is evaluated on the public Lung Nodule Analysis (LUNA16) challenge dataset. Our method is able to accurately detect lung nodules at high sensitivity and specificity and achieves $90.4\%$ sensitivity with 1/8 false positive per scan which outperforms the state-of-the-art results $15.6\%$.  

\end{abstract}

\section{Introduction}

Lung cancer is one of the leading cancer killers around the world which makes the study of lung cancer diagnosis eminently crucial. Computer-aided diagnosis (CAD) systems provide assistance for radiologists to accelerate the diagnosing process.  Many efforts \cite{Ding2017Accurate,khosravan2018s4nd,zhu2018deeplung} have been made for lung nodule detection by generalizing the recent powerful deep detection models in computer vision. Although these efforts made good progress in accurately detecting pulmonary nodules from CT scans, the false positive rate is still very high which limits the real application in routine clinical practice. For example, most of the previous work \cite{Ding2017Accurate,khosravan2018s4nd,zhu2018deeplung,wang2018automated} obtained less than $75\%$ sensitivity with 1/8 false positives per scan. To get sensitivity scores as high as $95.8\%$, these models would bear about eight false positives, which prevent their use in routine clinical practice.

We believe two main challenges prevent the existing models from accurate lung nodule detection. 1) Some normal tissues have similar morphological appearances as nodules in CT images which cause high false positives by wrongly detecting these tissues as nodules. 2) The high discrepancy of the volume between nodules and the whole CT scan may cause missing detection of real nodules. For example, the volume of a nodule with 10mm size in the diameter only occupied $0.059\%$ of the volume of whole CT scan (in average $213\times293$ pixels with $281$ slices). 
Furthermore, the size of pulmonary nodules can vary by as much as 10 times. For example, nodules in diameter can range from 3mm to 30mm in LUNA16 dataset. Therefore, it is particularly crucial for designing methods which can detect small volume nodules from large volume CT scans as well as to differentiate nodules from tissues with similar appearances in CT images. 

To address the above two challenges, this paper attempts to integrate the most recent progress in computer vision as well as the domain expert knowledge in medical imaging. Motivated by the-state-of the-art image detector using 2D Feature Pyramid Network (FPN) \cite{lin2017feature}, we develop a 3D feature pyramid network (3DFPN) for small nodule detection by appending the low-level high-resolution features to the high-level strong semantic features and a multi-scale feature prediction guarantees the wide-scale nodule detection. In addition, we carefully analyze the difference between nodules and the tissues where are wrongly detected as nodule candidates, and find that although they look similar on single CT slice, their spatial variances within continuous CT slices are distributed differently. This unique insight motivates us to design a novel refinement network, based on the location history image in the continous CT slices. The final model benefits from both the powerful deep network and medical imaging insights, and reduces a significant amount of the previous false detected nodule candidates. Our model achieves a $90.4\%$ sensitivity at 1/8 false positive per scan which significantly outperforms the state-of-the-art method $15.6\%$.

\section{Our Method}
\begin{figure}
    \includegraphics[width=118mm]{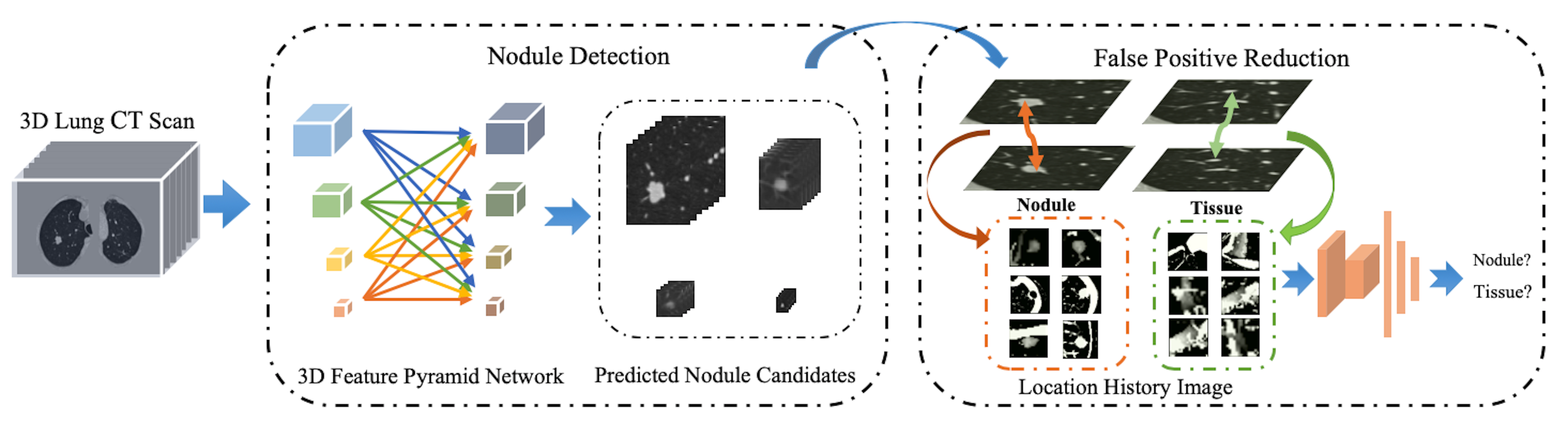}
    \caption{ The proposed 3DFPN-HS$^2$ framework of high sensitivity and specificity lung nodule detection by combining a 3D Feature Pyramid ConvNet (3DFPN) with an HS$^2$ network. A whole CT scan is fed into the 3DFPN to predict nodule candidates. For the detected nodule candidates, the HS$^2$ network eliminates the miss-predicted normal tissues based on the location variance within continuous CT slices. More detailed structure of the proposed 3DFPN network can be found in the supplementary document.
    } 
\vspace{-10pt}
\label{fig:system}
\end{figure}

As shown in Fig.\ref{fig:system}, the input of our lung nodule detection framework is a whole 3D volume of CT scan which is fed into a 3D Feature Pyramid ConvNet (3DFPN) to detect the 3D locations of nodule candidates. After the nodule candidates are detected,
we crop a 3D cube region centered with the candidate and develop a High Sensitivity and Specificity (HS$^2$) network to further recognize whether the detected nodule candidates are real nodules or false detected tissues which effectively reduces false positives. In this framework, the 3DFPN benefits from the progress of state-of-the-art deep learning, and the HS$^2$ network benefits from the insight of medical images. We will discuss them respectively. 

\subsection{3D Feature Pyramid ConvNet}
The recent progress in computer vision suggests feature pyramid networks (FPNs) are good at detect objects at different scales \cite{lin2017feature}. However, traditional FPNs are designed for 2D images. Here, we propose a 3DFPN network to detect 3D locations of lung nodules from 3D CT volumetric scans. Different than~\cite{lin2017feature} which only concatenates the upper-level features in feature pyramid, we use a dense pyramid network to integrate both the low-level high-resolution features as well as high-level high-semantic features, which enriches the location details and strong semantics for nodule detection. Table~\ref{fpns} highlights the main differences between 2DFPN and our 3DFPN. 

\vspace{-12pt}
\begin{table}[h!]
\caption{ Comparison between 2DFPN \cite{lin2017feature} and our proposed 3DFPN. We take 3D volume as input and the feature pyramid layers are integrated with lateral connections of all the high-level and low-level features.
}\label{fpns}
\resizebox{\textwidth}{!}{
\begin{tabular}{|l|c|c|c|c|c|c|c|}
\hline
Method  & \tabincell{c}{Input 3D\\volume} & \tabincell{c}{Lateral\\ connections} & \tabincell{c}{Integrate \\upper layer}  & \tabincell{c}{Integrate \\lower layer}   & \tabincell{c}{Upsample \\higher layer} &  \tabincell{c}{Downsample \\lower layer} \\
\hline
2DFPN\cite{lin2017feature}  & No & $\surd$ & $\surd$ &   &$\surd$ & \\
\hline
3DFPN   & $\surd$ & $\surd$ &$\surd$& $\surd$  &$\surd$&$\surd$\\
\hline
\end{tabular}}
\end{table}
\vspace{-12pt}

The bottom-up network extracts features from the convolution layer 2--5, refer as {C2, C3, C4, C5}, is followed by a convolution layer with kernel size $1\times1\times1$ to convert feature channels with the same number of channels. The feature pyramid network contains four layers, as {P2, P3, P4, P5}, which integrates the low-level features by a max pooling layer and the downsampled high-level features by deconvolution. 3DFPN predicts location with four parameters as $[x,y,z,d]$, where $[x,y]$ as the spatial coordinates at each CT slices, $z$ as CT slice number, and $d$ as nodule diameter and a confidence score for each candidate.

\begin{figure}[h!]
    \includegraphics[width=118mm]{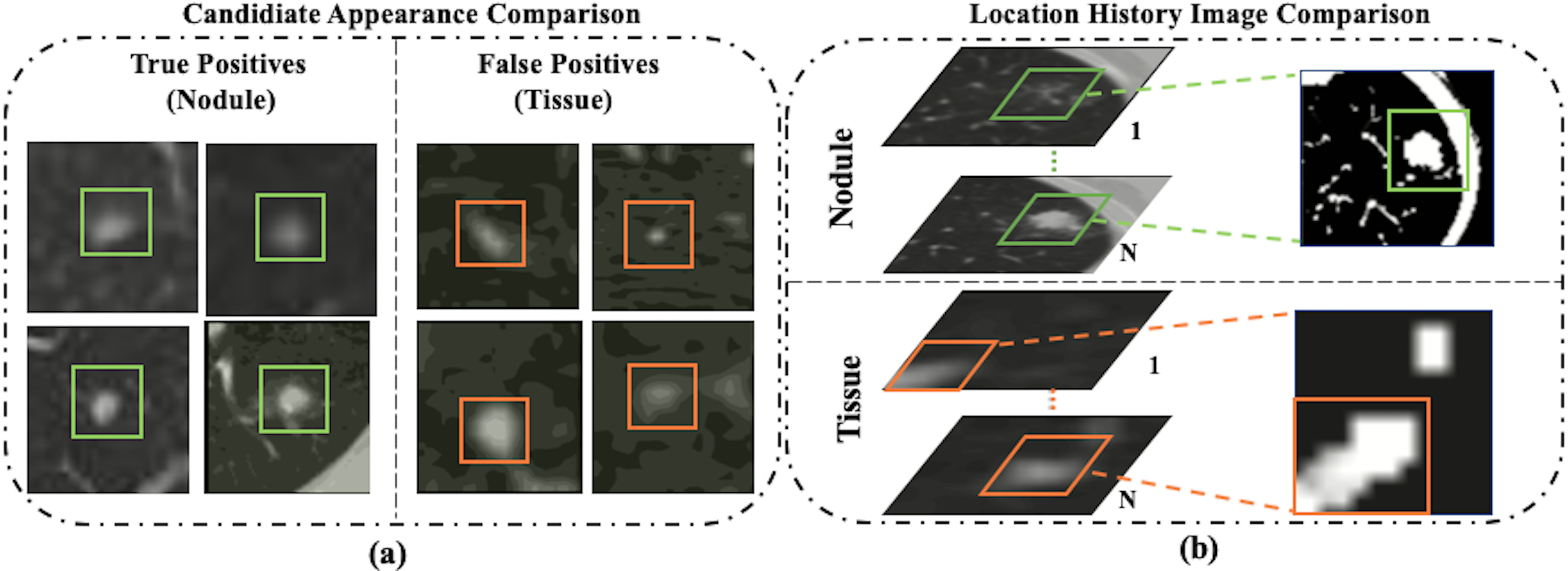}
    \caption{The proposed Location History Images (LHI) to distinguish tissues and nodules from the predicted nodule candidates. (a) Similar appearance of true nodules (green boxes) and false detected tissues (orange boxes). (b) The orientations of the location variances for nodules and tissues are presented differently in LHIs. True nodules generally have a circular region which indicates the spatial changes with either a brighter center (when nodule sizes in following CT slices are smaller) or a darker center (when nodule sizes in following CT slices are bigger). On the other hand, the location variance for false detected tissues usually tends to change in certain directions such as a gradually changed trajectory line. 
    } 
\vspace{-15pt}   
\label{fig:candiates}
\end{figure}

\subsection{HS$^2$ Network}

Due to the low resolution and the noise of CT images, as shown in Fig.~\ref{fig:candiates}(a), some tissues (orange boxes) appear to have similar features as real nodules (green boxes) which are very likely to be detected as nodule candidates. This leads to a large number of false positives. 
As shown in Table~\ref{nodulestats}, we further analyze 300 false positives predicted by the 3DFPN and observe that 241 False Positives (FPs) are caused by the high similarity of tissues (80.3\%), 33 of them are caused by inaccurate size detection, and 26 FPs are due to the inaccurate location detection.

\vspace{-15pt}
\begin{table}
\caption{Statistic Analysis for False Positive Nodule Candidates.
}\label{nodulestats}
\centering \resizebox{8cm}{!}{
\begin{tabular}{|l|c|c|c|}
\hline
& Tissue & Inaccurate Size & Inaccurate Location \\
\hline
Percentage& 80.3\% & 11\% & 8.7\%  \\
\hline
\end{tabular}}
\end{table}
\vspace{-15pt}

It is crucial to obtain a major difference to distinguish similar tissues from nodules for false positive reduction. By observing the continues slices, we discover that for tissues, the orientation of the location changes could be tracked in certain patterns, while the variance of true nodules tends to expand outside the contour or diminishing to the center at continuous CT slices. For instance, Fig.\ref{fig:candiates}(b) shows nodules and tissues in the appearance of continuous slides. For a gray-scale value of each pixel represents the closest change of the pixel in the region of nodules within a series of CT slices, with the gray value orthogonal to the movement, we obtain the location variance of the candidates.

Inspired by Motion History Image (MHI)~\cite{davis2001hierarchical}\cite{yang2012recognizing}, we define the Location History Image (LHI) as $f$. By given any pixel location $(x,y)$ on a CT slice $s$, $f(x,y,s)$ represents the intensity value of LHI within $(1, \tau)$ slice. The LHI is fed to a HS$^2$ (high sensitivity and specificity) network which is a feed-forward neural network with 2 convolution layers followed by 3 fully connected layers. The outputs of the HS$^2$ network are refined predicted labels of true nodules and tissues. Sensitivity is defined as a ratio of true positives over the total number of true positives and false negatives. Specificity is the ratio of true negatives over the total number of the true negatives and false positives.

The intensity of LHI is calculated as in Eq. (1):
\begin{equation}
 f(x,y,s)=\left\{
\begin{array}{lcl}
\tau     &      & {if~~\psi(x,y,s) = 1} \\
max(0,f(x,y,s-1)-1)    &      & {otherwise}
\end{array} \right. 
\end{equation}
where the update function $\psi(x,y,s)$ given by the spatial differentiation of the pixel intensity of two continuous CT slices. 
The algorithm has the following steps. 1) If $| I(x,y,s) - I(x,y,s-1)|$ is larger than a threshold, $\psi(x,y,s) = 1$, otherwise, $\psi(x,y,s) = 0$. 2) For the current slice, if $\psi(x,y,s) = 1$, $f = \tau$. Otherwise, if $f(x,y,s)$ is not zero, it is attenuated with a gradient of 1. If $f(x,y,s)$ equals zero, then remains as zero. 3) Repeat steps 1) and 2) until all the slices are processed.
Therefore, the location variance among continues CT slices and their change patterns can be effectively represented by our proposed LHIs.

\section{Experimental Results and Discussions}

\subsection{Dataset and Evaluation} The performance of the proposed framework is evaluated on the most popular LUNA16 challenge dataset~\cite{Aaa2017Validation} which consists of 1186 nodules in the size between $3-30$ mm from 888 CT scans and agreed by at least 3 out of 4 radiologists. It is divided into 10 subsets. In order to conduct a fair comparison with other methods, we follow the same process to conduct cross validations by using 9 subsets for training and the remaining 1 subset for testing, then obtain the final results by averaging the 10 experiments. Data augment is applied by flipping and resizing the CT scans. Same as other methods, the Free-Response Receiver Operating Characteristic (FROC) analysis \cite{Setio2016Pulmonary} and Competition Performance Metric (CPM) of detection sensitivity and the corresponding false positives at $1/8, 1/4, 1/2, 1, 2, 4, 8$ per scan are employed to measure the performance. The CPM score is calculated by the average of sensitivity for all the levels of false positives per scan.

\begin{table}
\caption{FROC Performance comparison with the state-of-the-arts: sensitivity (recall) and the corresponding false positives at $1/8, 1/4, 1/2, 1, 2, 4, 8$ per scan. Our 3DFPN-HS$^2$ method achieves the best performance (with $>90\%$ sensitivity) at all false positive levels and significantly outperforms others especially at the low false positive levels ($1/8$ and $1/4$).
}\label{tab1e:overall}
\centering
\begin{tabular}{|l|c|c|c|c|c|c|c|c|}
\hline
\centering {\bfseries Methods} &  {\bfseries1/8 } & {\bfseries1/4 }& {\bfseries1/2 }&{\bfseries1 }& {\bfseries2} & {\bfseries4} & {\bfseries8} & {\bfseries CPM score}\\
\hline
 Dou et al.\cite{Dou2017Automated} & 0.659 & 0.745 & 0.819 & 0.865 & 0.906 &0.933&	0.946&	0.839\\
 Zhu et al.\cite{zhu2018deeplung}& 0.692 & 0.769 & 0.824 & 0.865 & 0.893 & 0.917& 0.933 & 0.842\\
 Wang et al.\cite{wang2018automated}& 0.676 & 0.776 & 0.879 & 0.949 & 0.958 & 0.958& 0.958 & 0.878\\
 Ding et al.\cite{Ding2017Accurate}& 0.748 &0.853 &0.887& 0.922& 0.938& 0.944 &0.946& 0.891\\
 Khosravan et al.\cite{khosravan2018s4nd} & 0.709& 	0.836&	0.921& 0.953& 0.953& 0.953& 0.953& 0.897\\
 {\bfseries 3DFPN (Ours)} & {\bfseries 0.848} & {\bfseries 0.876} & {\bfseries 0.905} & {\bfseries 0.933}& {\bfseries0.943}&{\bfseries0.957}&{\bfseries0.970}& {\bfseries0.919}\\
 {\bfseries 3DFPN-HS$^2$ (Ours)} & {\bfseries 0.904} & {\bfseries 0.914} &  {\bfseries0.933} & {\bfseries0.957} &  {\bfseries0.971} &{\bfseries0.971}&{\bfseries0.971}& {\bfseries0.952}\\
\hline
\end{tabular}
\vspace{-12pt}
\end{table}

\subsection{Experimental Results}
\textbf{Experimental Settings.}
The framework takes the whole CT scan as input, while volume at $96\times96\times96$ pixels is selected by a sliding window method as the input of the 3DFPN network. This size is selected based on experiments which is bigger enough to contain the whole nodule even when it is with the largest size (about 30 mm). The anchor sizes employed in our 3DFPN to obtain the candidate regions from feature maps are [3$^3$, 5$^3$, 10$^3$, 15$^3$, 20$^3$, 25$^3$, 30$^3$] pixels. For all the anchors, the corresponding regions obtained from all the 3D feature map levels are gathered to predict the position of nodules.

In the training phase, the regions with an Intersection-over-Union (IoU) threshold to the ground-truth regions less than $0.02$ are referred to negative samples and greater than 0.4 are positive samples. The samples in between are ignored to avoid the positive and negative samples similarity. A classification layer is used to predict a confidence score for the candidate class and a region regression layer is applied to learn the offset between the position of region proposals and the ground-truth. We adopt Smooth $L1$ loss \cite{girshick2015fast} and binary cross entropy loss (BCE-loss) for location regression and classification score respectively. 
In the testing, for each region proposal, a confidence score is calculated by the classification layer. The proposals with a probability larger than 0.1 are chosen as nodule candidates. Non-maximum suppression is further applied to eliminate the multiple predicted candidates for one nodule.

The two convolution layers of the HS$^2$ network are set to $(1, 30)$, $(30, 50)$ dimensions and followed by three fully connected layers with the channel sizes of $(2048, 1024, 512)$. The cross entropy loss is applied for classification during the training. Image patches aligned with each predicted nodule candidate region but with twice size (in both x and y directions) are selected from $11$ continuous CT slices ($5$ slices before and after the current slice of nodule candidate respectively). The LHI of these patches is extracted and resized to $48\times48$ pixels as the input of the HS$^2$ network.

In the training, the learning rate starts from $0.01$ and decreases to $1/10$ for every $500$ epochs. Total of $2,000$ epochs is conducted for the framework. The average prediction time for a whole CT scan is about $0.53$ min/scan on a server with one GeForce GTX 1080 GPU using Pytorch 2.7.

\textbf{Comparison with Other Methods.} Table~\ref{tab1e:overall} shows the FROC evaluation results with ($1/8,1/4,1/2,1,2,4,8$) false positive levels of our proposed method compared with state-of-the-art methods. The highlighted numbers in the table indicate the best performance within each column. All the methods are tested on LUNA16 dataset followed the same FROC evaluation. As shown in the table, our framework outperforms $5.5\%$ average sensitivity than the best result of other methods. In addition, the proposed framework achieves the best performance at every FP level. As previously mentioned, the CAD system is not only required a high sensitivity, but also a high specificity.  Table~\ref{tab1e:overall} demonstrates that the false positives are greatly reduced by the proposed HS$^2$ network. 3DFPN-HS$^2$ obtains a highest $97.14\%$ sensitivity at 2 FPs per scan. In addition, for the FP of 1/8, 1/4, and 1/2 per scan, the proposed framework still remains a high sensitivity above $90\%$. The experimental results show that 3DFPN-HS$^2$ reaches a state-of-the-art performance for high sensitivity and specificity lung nodule detection.

\begin{figure}[tt]

    \includegraphics[width=118mm]{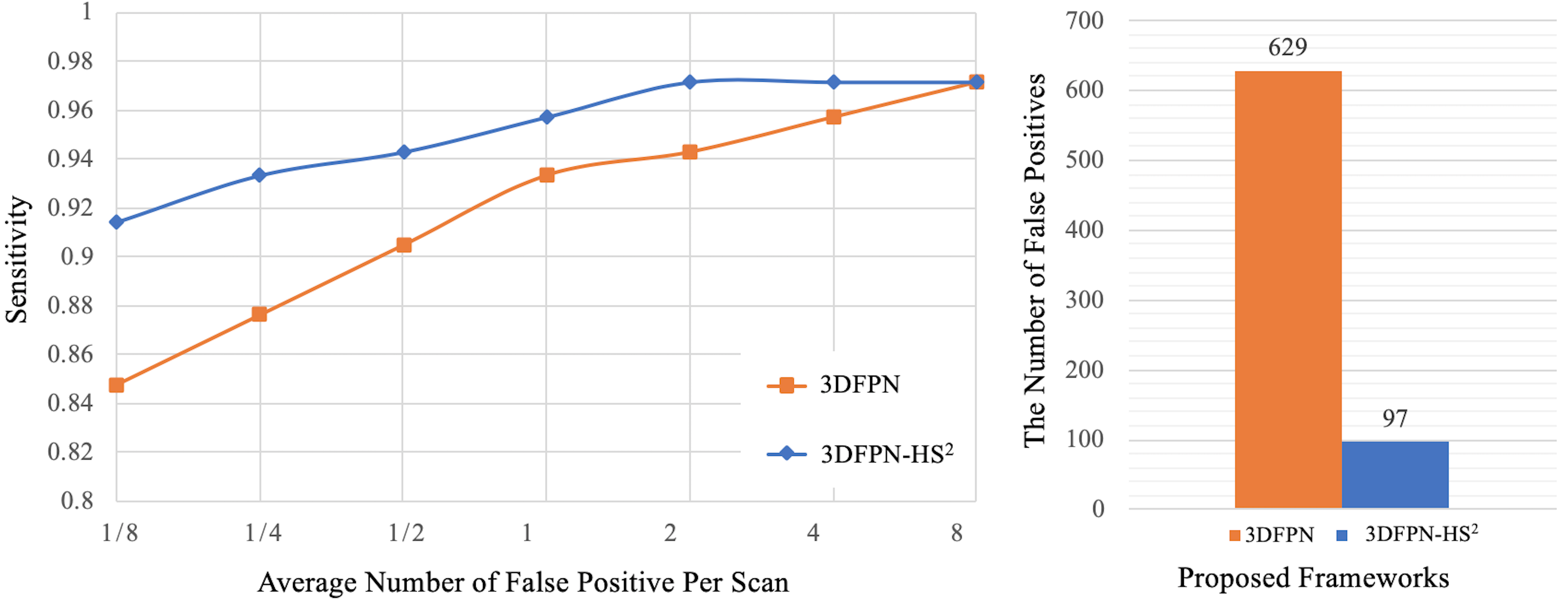}
    \caption{Left: Comparison between the proposed 3DFPN and 3DFPN-HS$^2$ (with High Sensitivity and Specificity network for false positive reduction.) 3DFPN-HS$^2$ greatly improves the performance of the 3DFPN at all the FP levels. Right: The number of false positives is reduced from $629$ to $97$ for a total of $88$ CT scans with the confidence score above $0$ after the HS$^2$ network is applied. More visualized detection results are provided in the supplementary document.
    } 
     \vspace{-0.5cm}
\label{fig:Result1}
\end{figure}

\textbf{Effectiveness of HS$^2$ for FP Reduction.} Two experiments are conducted to demonstrate the advantages of HS$^2$ network. As shown in Fig.~\ref{fig:Result1}(a), compared with 3DFPN without the HS$^2$ network, the result of 3DFPN-HS$^2$ with the false positive reduction is increased more than $5\%$ at 1/8 FP level. In addition, the numbers of FPs with (blue bar) and without (orange bar) HS$^2$ network for all the predicted nodule candidates from a total of 88 CT scans (subset 9) are further compared in  Fig.~\ref{fig:Result1}(b). By applying HS$^2$, the 3DFPN-HS$^2$ is able to distinguish the falsely detected tissues from true nodules, therefore significantly reduces FPs by $84.5\%$. It is worth noting that our proposed 3DFPN without HS$^2$ network still achieves $97\%$ at $8$ FPs per scan and $91.9\%$ CPM, which surpasses other state-of-the-art methods (see Table~\ref{tab1e:overall}.)

\section{Conclusion}
In this paper, we have proposed an effective framework 3DFPN-HS$^2$ by employing a 3D feature pyramid network with local and global feature enrichment for small volume and multi-scale nodule detection. HS$^2$ network is introduced to reduce false positives based on the different patterns of location variance for nodules and tissues in continuous CT slices. The proposed framework significantly outperforms the state-of-the-art methods and has achieved high sensitivity and specificity which has a great potential in routine clinical practice. 

\vspace{10pt}
\noindent
\textbf{Acknowledgements.} This material is based upon work supported by the National Science Foundation under award number IIS-1400802 and Memorial  Sloan-Kettering Cancer Center Support Grant/Core Grant P30 CA008748. Oguz Akin, MD serves as a scientific advisor for Ezra AI, Inc., which is developing artificial intelligence algorithms and software unrelated to the research being reported.
\bibliography{sample.bib}
\bibliographystyle{splncs04}
\end{document}